\documentclass[onecolumn, superscriptaddress, amsmath, amssymb, aps, prx]{revtex4-2}

\usepackage{graphicx}
\usepackage{dcolumn}
\usepackage{bm}
\usepackage{amsmath}
\usepackage{comment}
\usepackage{makecell}
\usepackage{rotating}   
\usepackage{array}      
\usepackage{tabularx}   

\begin{document}

\preprint{APS/123-QED}

\title{Spatial coherence enabled sensorless adaptive optical imaging}

\author{Pranay Mohta}
\affiliation{Department of Physics, Indian Institute of Technology Kanpur, Kanpur, UP 208016, India}

\author{Shaurya Aarav}
\affiliation{Sorbonne Universit\'e, CNRS, Institut des NanoSciences de Paris, INSP, F-75005 Paris, France}
  
\author{Hugo Defienne}
\affiliation{Sorbonne Universit\'e, CNRS, Institut des NanoSciences de Paris, INSP, F-75005 Paris, France}

\author{Anand K. Jha}
\affiliation{Department of Physics, Indian Institute of Technology Kanpur, Kanpur, UP 208016, India}


\begin{abstract}
Optical aberrations degrade imaging performance in label-free microscopy, where the absence of a guide star often necessitates sensorless adaptive optics (AO). Conventional sensorless AO approaches rely on image-quality metrics whose optimal choice depends on both the specimen and the imaging modality. Here, we present a guide-star-free AO framework based on the spatial coherence properties of spatially incoherent light. The proposed method exploits aberration-induced broadening of the measured spatial correlation distribution as the feedback signal for aberration correction. Experiments in a conventional bright-field imaging system using standard LED illumination demonstrate successful correction of phase aberrations. Furthermore, the approach remains effective even in the presence of spatially structured background noise. These results establish spatial coherence measurements as an effective feedback mechanism for sensorless AO and indicate that the correlation-based feedback principle employed in quantum-assisted AO can likewise be realized using the spatial correlations of incoherent light.
\end{abstract}

\maketitle


\section{\label{sec:level1}Introduction}

Optical microscopy uses light to resolve structures too small to be seen with the naked eye. Due to its non-invasive nature and high spatial resolution, it has become a cornerstone technique in biomedical research. Over the past few decades, the field has seen rapid advancements, leading to significant improvements in imaging capabilities. Nevertheless, key performance parameters—such as resolution, signal-to-ratio (SNR) and imaging depth—remain fundamentally limited by optical aberrations \cite{hampson2021nrmp, booth2015microscopy}. These aberrations arise from imperfections in optical components and system alignment, as well as from refractive index variations within heterogeneous samples.

To mitigate these distortions, adaptive optics (AO)—originally developed for ground-based astronomy~\cite{beckers1993araa, hampson2021nrmp, bolbasova2022aoo, tyson2022principles}—has been incorporated into modern microscopy to quantify and correct optical aberrations~\cite{booth2015microscopy, zhang2023boe, ji2017nm}. An AO system typically involves two key steps: aberration estimation and its subsequent correction. In conventional direct-sensing approaches, a localized point-like light source within the sample, referred to as a guide star, serves as a reference. As light from the guide star propagates through the sample and imaging system, its wavefront accumulates distortions, which are measured using a Shack–Hartmann wavefront sensor. The estimated aberration is then compensated by applying an appropriate phase correction, typically using a deformable mirror or a spatial light modulator (SLM). A number of other techniques for wavefront sensing have also been used like interferometric focus sensing~\cite{papadopoulos2016np}, machine learning~\cite{guo2022oea} and digital holography~\cite{upatnieks1966ao, kim2012ol, liu2013ao}.

In scenarios where a physical guide star cannot be inserted or identified—a frequent challenge in label-free microscopy—direct wavefront sensing becomes impractical. To address this, various sensorless adaptive optics strategies have been developed~\cite{muller1974josaa, zhang2023boe, hampson2020, booth2006oe}. Rather than measuring the wavefront directly, a conventional class of these techniques utilizes an iterative optimization framework centered on an image-quality based metric. By modulating a wavefront-shaping element, the system seeks to maximize a specific quality metric, such as spatial frequency distribution~\cite{debarre2007oe}, image sharpness~\cite{fienup2003josaa, murray2005}, contrast~\cite{zhou2015boe} or integrated signal intensity~\cite{debarre2009ol}. In recent years, several label-free microscopy techniques have incorporated AO for aberration correction~\cite{debarre2007oe, shu2022px, kam2001pnas, jian2014boe}. However, the choice of metric is closely tied to both the imaging mechanism and the properties of the sample.

Beyond image-based metrics, other sensorless AO frameworks have emerged. One such approach is quantum-assisted AO (QAO)~\cite{cameron2024sc}, which leverages the spatial correlations of entangled photons. While QAO operates independently of the sample and imaging modality without requiring a guide star, its practical utility is hampered by the low generation efficiency of entangled photons and the inherent complexity of time-consuming coincidence detection. More recently, an alternative based on intensity-intensity correlations using even-symmetrical thermal light was proposed~\cite{he2025aplph}. However, this technique requires highly specialized illumination, rendering it incompatible with standard, widely used incoherent light sources.

In this paper, we present a spatial-coherence-based sensorless AO framework that retrieves aberration information in the absence of a guide star while operating under standard spatially incoherent illumination. Under ideal conditions, light from a spatially incoherent source is uncorrelated across different spatial locations. Optical aberrations modify this correlation state, causing the measured spatial correlation distribution to broaden. By quantifying this aberration-induced broadening, the proposed method provides a coherence-based feedback signal for iterative aberration correction without relying explicitly on object intensity features. We experimentally validate the approach by imaging different samples through a standard incoherent imaging system in the presence of aberrations. We further demonstrate robust aberration correction in the presence of both spatially uniform and structured background noise. Our results demonstrate that spatial coherence measurements can provide an effective feedback observable for sensorless AO and show that the correlation-based feedback principle employed in QAO can also be realized using the spatial correlations of incoherent light.
\begin{figure}[b]
\centering
\includegraphics[scale=0.70]{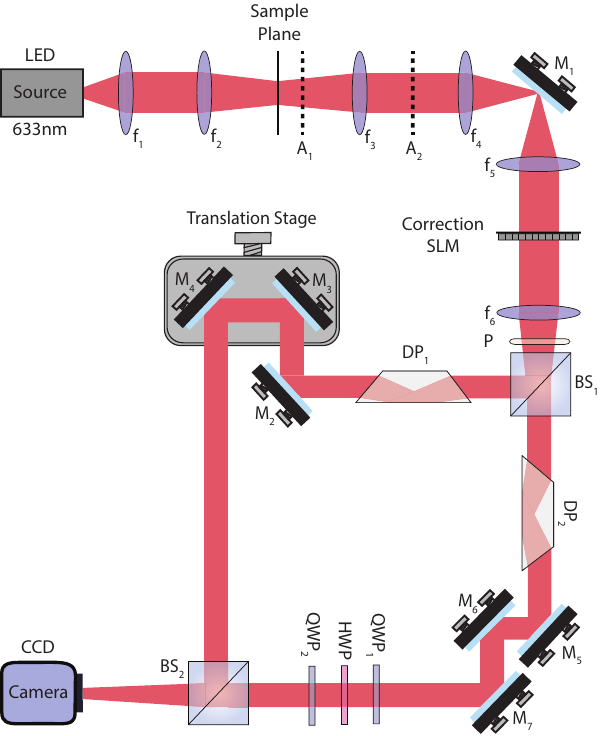}
\caption{Experimental setup of the spatial coherence-based adaptive optics system. An LED source provides spatially incoherent illumination, which is relayed to the sample plane through the $f_1$-$f_2$ 4f imaging system. The sample is subsequently imaged onto the camera using two successive 4f systems, $f_3$-$f_4$ and $f_5$-$f_6$. Aberrations can be introduced at planes A1 and A2 to emulate specimen- and system-induced distortions, respectively. A spatial light modulator (SLM) positioned at the Fourier plane applies aberration correction. A wavefront-inversion interferometer placed before the camera measures the cross-spectral density (CSD) at the image plane. Image acquisition is performed by blocking one interferometer arm and recording the resulting intensity distribution. L: thin lenses, BS: beam splitter, DP: dove prism, P: polariser, QWP: quarter-wave plate, HWP: half-wave plate and TS: linear translation stage.}
\label{cao_setup}
\end{figure}
\begin{figure*}[t]
\centering
\includegraphics[scale=0.66]{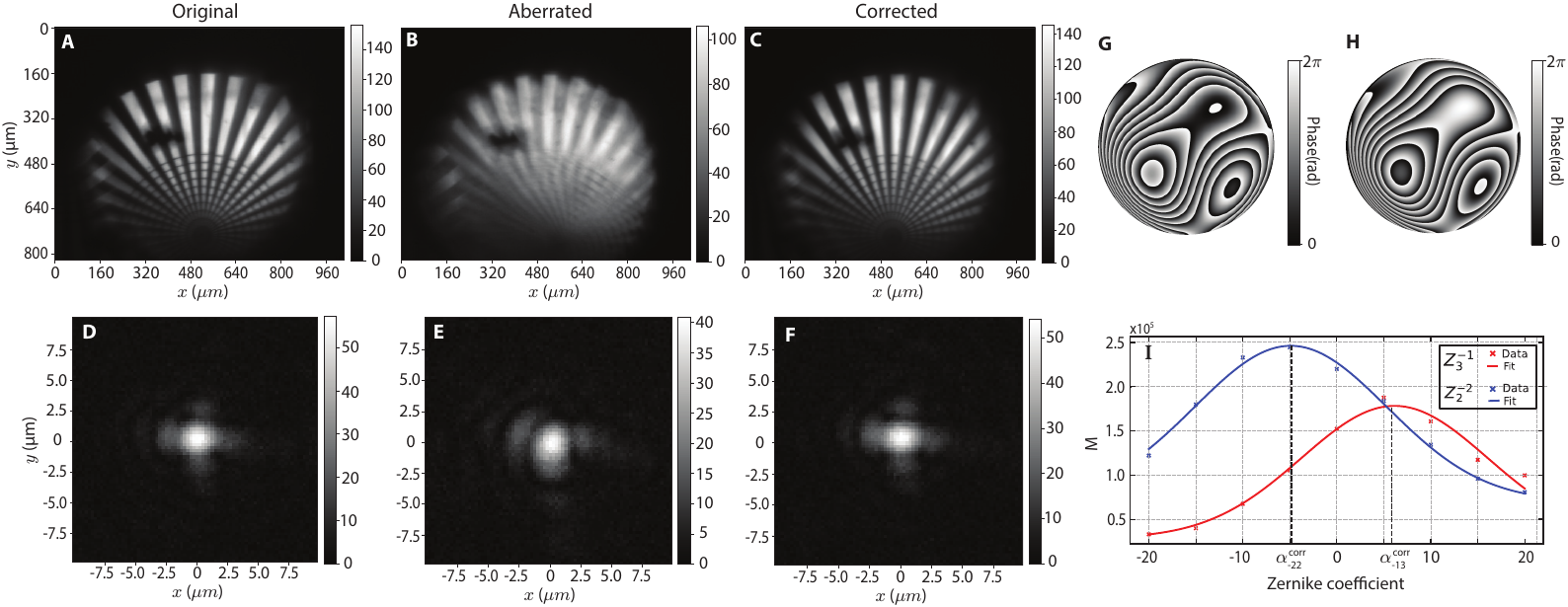}
\caption{Correction of SLM-induced random phase aberrations. (A) Aberration-free image of the resolution target. (B) Aberrated image obtained after applying a low-spatial-frequency random phase mask at plane A2. (C) Corrected image after optimization. (D)-(F) Corresponding measured CSD distributions for the aberration-free, aberrated, and corrected cases, respectively. (G) Phase profile of the induced aberration. (H) Optimized correction phase applied to the correction SLM. (I) Variation of the coherence-sharpness metric $M$ with bias amplitude $\alpha_{mn}$ for the Zernike modes $Z_3^{-1}$ and $Z_2^{-2}$. Gaussian fitting is used to determine the optimal correction coefficients $\alpha^{\mathrm{corr}}_{-1,3}=6.14$ and $\alpha^{\mathrm{corr}}_{-2,2}=-4.42$. The SSIM improves from $0.69$ in the aberrated case to $0.90$ after correction.}
\label{cao_result_1}
\end{figure*}

\section{\label{sec:level1-results}Concept}
Spatial coherence of an optical field describes the correlation of the electric field at two spatial locations. It is quantified by the cross-spectral density function (CSD). The electric field vibrations are perfectly correlated across all spatial points for a spatially coherent source, whereas for a spatially incoherent source, they are completely uncorrelated. We consider imaging with incoherent light. For an ideal, quasi-monochromatic incoherent source the CSD is of the form,
\begin{equation}
    W(\mathbf{r}_1, \mathbf{r}_2) = I(\mathbf{r}_1)\delta(\mathbf{r}_1 - \mathbf{r}_2)
\end{equation}
where $I(\mathbf{r})$ is the spectral density of the source and $\delta$ denotes the Dirac delta function. To experimentally access this correlation information, we employ a wavefront-inversion interferometer \cite{murty1964josa,wessely1970josaa,breckinridge1972ao,bhattacharjee2018apl,mohta2025jop}. This device interferes the optical field with a spatially inverted version of itself, effectively probing the CSD at coordinate pairs, $\mathbf{r}_1 = \mathbf{r}$ and $\mathbf{r}_2 = -\mathbf{r}$, as described in Methods B. The measured CSD is given by,
\begin{equation}
    W_{\mathrm{img}}(\mathbf{r}, -\mathbf{r}) = \int I_0(\bm{\rho}) h(\mathbf{r} - \bm{\rho}) h^*(-\mathbf{r} - \bm{\rho}) \, d\bm{\rho}.
\end{equation}
where $h(\mathbf{r})$ denotes the coherent point spread function (PSF) of the imaging system. In the absence of aberration, the measured CSD $W(\mathbf{r}, -\mathbf{r})$ is narrowly localized around the optical axis ($\mathbf{r} = 0$), manifesting as a singular, sharp correlation peak. The presence of optical aberrations reduces the spatial localization of the measured CSD. Inspired by the conventional image-sharpness metric, we define a coherence-sharpness metric to characterize the localization of the measured CSD distribution,
\begin{equation}
M=\iint |W(\mathbf{r},-\mathbf{r})|^2\,d\mathbf{r},
\end{equation}
It serves as a measure of the correlation localization and reaches its maximum when no aberration is present. We employ this optimal value as a feedback signal within a modal-based adaptive optics algorithm to iteratively correct aberrations. Theoretical analysis and numerical simulations supporting the proposed metric are presented in Methods A.

\section{\label{sec:level1-results}Results}

Figure~\ref{cao_setup} illustrates the experimental setup used in this work. A spatially incoherent LED source~\cite{tziraki2000apb,diLorenzoPires2010ol,torcalMilla2022optik,deng2017scr} illuminates the sample, while aberrations can be introduced either by the specimen or within the imaging system. A spatial light modulator (SLM) positioned at the Fourier plane is used to compensate these aberrations. The corrected optical field is then analyzed using a wavefront-inversion interferometer, which measures the CSD at the image plane~\cite{mohta2025jop}. For intensity imaging, one interferometer arm is blocked and the image is recorded directly.

To demonstrate the proposed approach, a resolution test target was positioned at the sample plane, while optical aberrations were introduced using either a spatial light modulator (SLM) or a polydimethylsiloxane (PDMS) sheet. For aberration correction, we implemented a modal-based AO framework using coherence sharpness $M$ as the metric for iterative optimization. This process involves the sequential application of Zernike polynomial modes $Z_n^m$ to the correction SLM, covering all radial orders $n \leq 4$ and azimuthal indices $|m| \leq n$, while omitting the piston, tip, and tilt. For each specific mode, we recorded a series of CSD measurements under varying bias amplitudes $\alpha_{mn}$. The phase profile of the SLM was modulated according to: $\phi_{nm}=\phi^{\mathrm{corr}}_{nm-1}+\alpha_{mn}Z_n^m$.
where $\phi^{\mathrm{corr}}_{nm-1}$ represents the optimized phase derived from the preceding Zernike mode. By plotting the coherence-sharpness metric M as a function of the applied bias amplitude, we obtained an optimization curve. A Gaussian fit was then performed to determine the location of the peak response, from which the optimal correction coefficient was extracted and used to update the cumulative phase pattern $\phi^{\mathrm{corr}}_{nm}$.
Through successive optimization cycles across the defined modal basis, low-order aberrations were systematically minimized. This restoration led to a significant sharpening of the CSD peak and the subsequent retrieval of a high-fidelity image of the specimen. Such iterative mode-wise phase modulation is standard in classical modal AO~\cite{booth2006oe}.


%
\begin{figure*}[t]
\centering
\includegraphics[scale=0.74]{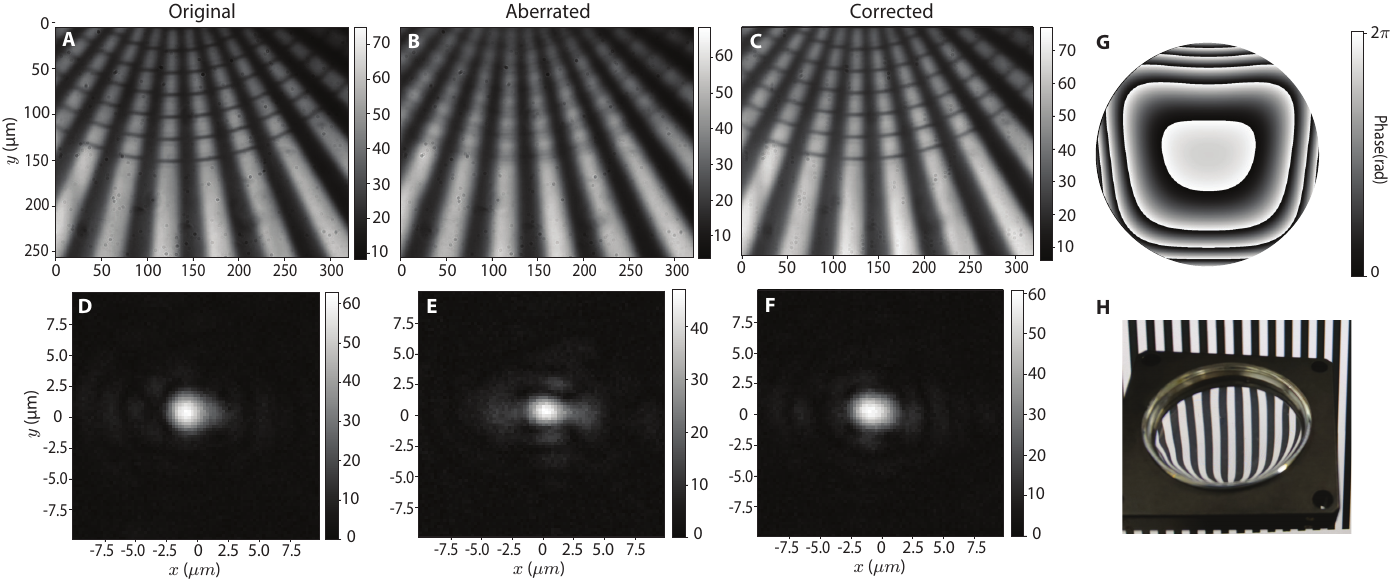}
\caption{Compensation of aberrations introduced by a physical translucent medium. A 1-cm-thick PDMS layer positioned at plane A1 introduces refractive-index-induced phase distortions. (A) Aberration-free image. (B) Image degraded by the PDMS layer. (C) Corrected image after optimization. (D)-(F) Corresponding measured CSD distributions for the aberration-free, aberrated, and corrected conditions, respectively. (G) Optimized correction phase used to compensate the PDMS-induced aberration. (H) Photograph of the 1-cm-thick PDMS layer used.
}
\label{cao_result_2}
\end{figure*}

To validate the performance of the proposed coherence-based adaptive optics framework, we conducted a series of experiments. We first examined the system’s response to phase distortions introduced via the SLM at the Fourier plane (plane A2). Figure~\ref{cao_result_1} summarizes the results of an experiment where a low-frequency random phase mask—constructed from a weighted combination of Zernike polynomials—was applied to the system. In the aberration-free state (Fig.~\ref{cao_result_1}A), the sample image is well-resolved and corresponds to a sharp, highly localized CSD profile (Fig.~\ref{cao_result_1}D). Upon introduction of aberration, the imaging fidelity is severely compromised, resulting in a blurred sample image (Fig.~\ref{cao_result_1}B) and a spatially distorted CSD (Fig.~\ref{cao_result_1}E).
The iterative AO correction successfully restores the field's coherence, returning the CSD to its sharp state (Fig.~\ref{cao_result_1}F). This is evident in the corrected image (Fig.~\ref{cao_result_1}C), which shows a stark improvement in clarity compared to the aberrated case. To quantitatively evaluate this enhancement, we utilized the Structural Similarity Index Measure (SSIM), with the aberration-free image (Fig.~\ref{cao_result_1}A) serving as the ground truth. SSIM improved from a baseline of 0.69 for the uncorrected state to 0.90 after optimization of the AO. The phase profiles corresponding to the induced distortion and the applied correction are shown in Figs.~\ref{cao_result_1}G and~\ref{cao_result_1}H, respectively. Furthermore, the optimization profile for two representative modes ($Z_3^{-1}$ and $Z_2^{-2}$) is plotted in Fig.~\ref{cao_result_1}I. This figure illustrates the plot of the metric $M$ with the bias amplitude $\alpha_{mn}$, where the peak locations $\alpha^{\mathrm{corr}}_{-1,3}$ and $\alpha^{\mathrm{corr}}_{-2,2}$—extracted from Gaussian fitting—identify the optimal correction coefficients for the respective modes.


We also test our scheme with a physical aberrating medium. A 1-cm-thick PDMS layer was used as a physical aberrating medium and positioned at plane A1, approximately 2 cm from the sample. Refractive index inhomogeneities within the PDMS introduced substantial phase distortions, resulting in image degradation (Fig.~\ref{cao_result_2}B) and broadening of the measured CSD distribution. Application of the proposed AO scheme compensated for these aberrations and restored image quality, as shown in Fig.~\ref{cao_result_2}C. Correspondingly, the CSD recovered its localized profile following correction (Fig.~\ref{cao_result_2}F). The optimized phase pattern used to compensate the PDMS-induced aberration is presented in Fig.~\ref{cao_result_2}G. Quantitatively, the SSIM improved from 0.75 for the aberrated image to 0.89 after correction.

\begin{figure*}[t]
\centering
\includegraphics[scale=0.68]{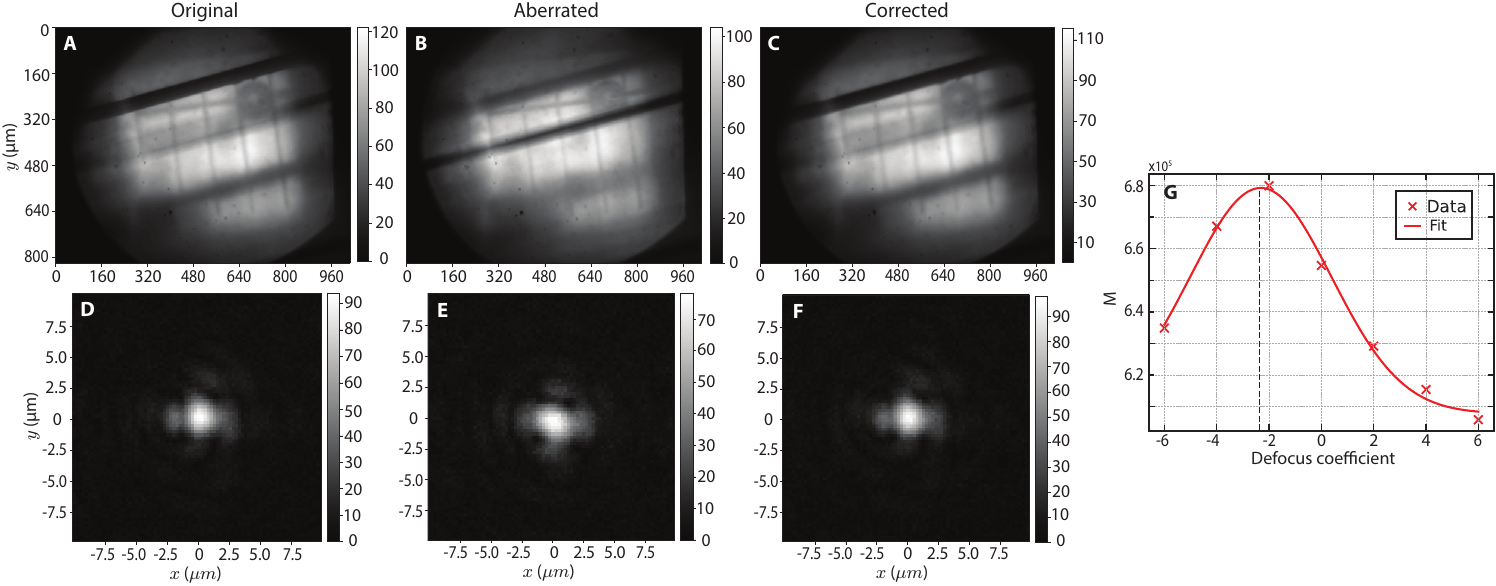}
\caption{Defocus correction in an axially extended object. The object consists of three human hair strands separated axially by approximately 4~mm. (A) Diffraction-limited image with the top strand in focus. (B) Image after introducing defocus aberration $(\alpha^{\mathrm{abrr}}_{0,2}=2)$, causing the focal plane to shift to the middle strand. (C) Corrected image after optimization, restoring focus to the original top strand. (D)-(F) Corresponding measured CSD distributions for the diffraction-limited, aberrated, and corrected conditions, respectively. (G) Variation of the coherence-sharpness metric $M$ as a function of the defocus bias amplitude. Gaussian fitting yields the optimal correction coefficient $\alpha^{\mathrm{corr}}_{0,2}=-2.31$. The SSIM improves from $0.76$ in the aberrated state to $0.92$ after correction.}
\label{cao_result_3}
\end{figure*}

As previously reported in QAO~\cite{cameron2024sc}, intensity-based metrics—such as power-in-the-bucket (PIB), contrast, and low-frequency spectral power—can introduce systematic errors when correcting for defocus aberration in 3D specimens. Specifically, these metrics may optimize for the wrong focal plane within a thick sample. We next examine the performance of the proposed coherence-based optimization for an axially extended object consisting of three human hairs separated by approximately 4 mm. In the initial unaberrated state, the top hair strand is in focus (Fig. \ref{cao_result_3}A). We then induced a defocus aberration ($\alpha^{\mathrm{abrr}}_{0,2} = 2$) via a second SLM at plane A2, which shifted the focal plane to the middle hair strand (Fig.~\ref{cao_result_3}B) and consequently reduced the CSD sharpness metric (Fig.~\ref{cao_result_3}E). Our optimization scheme successfully restored the CSD localization (Fig.~\ref{cao_result_3}F), returning the focus to the original top hair strand (Fig.~\ref{cao_result_3}D). The algorithm converged to a correction coefficient of $\alpha^{\mathrm{corr}}_{0,2} = -2.31$. Quantitative evaluation shows an increase in SSIM from 0.76 to 0.92. 

%

We next investigate the robustness of the proposed framework in the presence of background noise. Since conventional image-based AO derives its feedback directly from the recorded image intensity, background signals can lead to unreliable optimization, particularly when the background is spatially structured or comparable in strength to the object signal. In contrast, the proposed framework remains resilient to both uniform and spatially structured background noise. This robustness originates at the measurement stage rather than the optimization stage. As demonstrated in~\cite{mohta2026prap}, when the object illumination possesses a longer temporal coherence length than the background, the background contribution to the CSD can be significantly suppressed through temporal coherence filtering. By choosing the interferometer path-length difference to exceed the temporal coherence length of the background while remaining shorter than that of the object, interference from the background is strongly suppressed. Consequently, the measured CSD is dominated by the object field, whereas the background contributes negligibly to the correlation measurement. This enables effective signal-to-noise enhancement even when the object and background fields have substantial spectral overlap, where conventional spectral filtering becomes ineffective.

\begin{figure*}[t]
\centering
\includegraphics[scale=0.7]{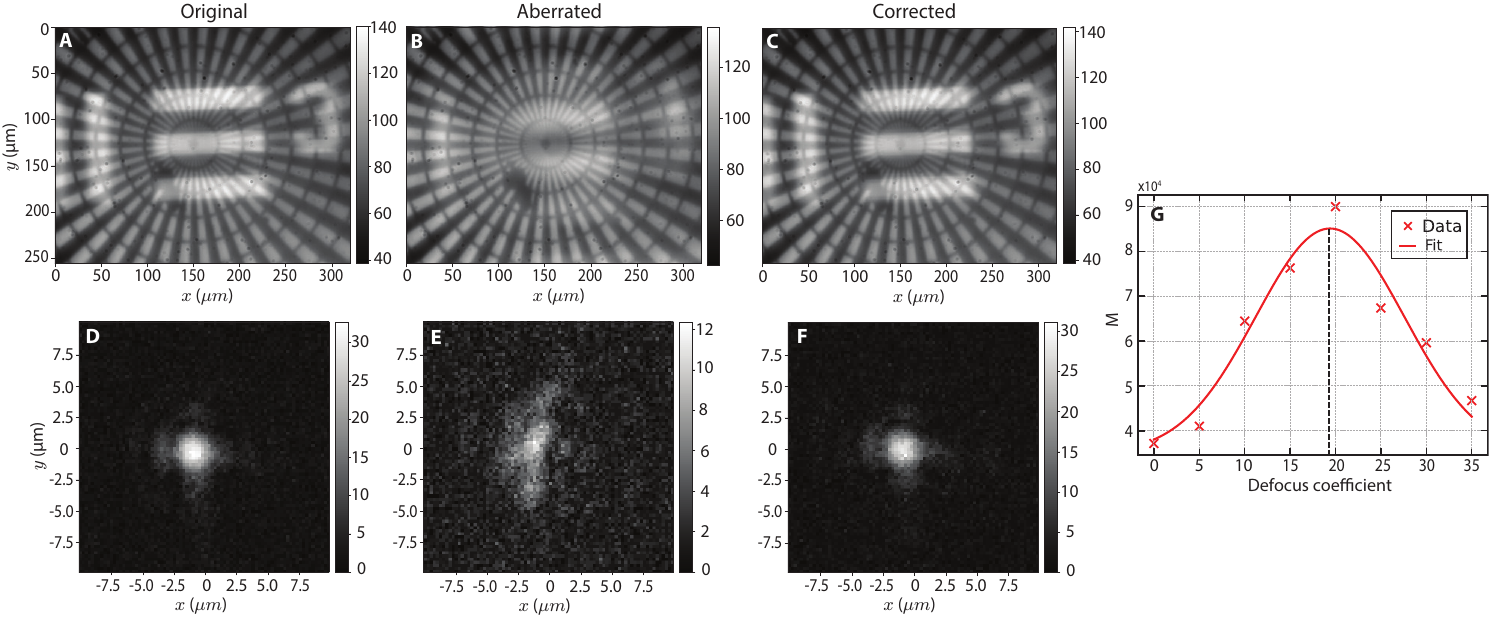}
\caption{Noise-resilient aberration correction. The sample is obscured by a structured background (Siemens star pattern) with a signal-to-noise ratio of $1:5$ based on total power. Temporal coherence filtering suppresses the background contribution in the measured CSD. (A) Sample image in the presence of structured background noise. (B) Image after introducing severe defocus aberration $(\alpha^{\mathrm{abrr}}_{0,2}=-20)$. (C) Corrected image after optimization. (D)-(F) Corresponding measured CSD distributions for the initial, aberrated, and corrected conditions, respectively. Despite the high-noise environment, the optimization restores both the localized CSD peak and the resolved sample image. (G) Optimization curve of the coherence-sharpness metric $M$ as a function of defocus bias amplitude. The optimal correction coefficient obtained is $\alpha^{\mathrm{corr}}_{0,2}=19.39$.}
\label{cao_result_4}
\end{figure*}

Experimentally, background noise was introduced at the first beam splitter of the interferometer, allowing it to propagate with the object field throughout the interferometer.  To demonstrate this capability, the sample was obscured by a spatially structured background consisting of a Siemens star pattern, as shown in Fig.~\ref{cao_result_4}A. Although the recorded intensity image contains contributions from both the object and the background, the corresponding CSD (Fig.~\ref{cao_result_4}D) primarily represents the object due to temporal coherence filtering. Introducing a defocus aberration ($\alpha^{\mathrm{abrr}}_{0,2}=-20$) caused the object to become completely obscured in the recorded image (Fig.~\ref{cao_result_4}B), while simultaneously broadening the measured CSD (Fig.~\ref{cao_result_4}E). Despite the background noise, the proposed AO framework successfully restored the localized CSD (Fig.~\ref{cao_result_4}F) and the corresponding corrected image (Fig.~\ref{cao_result_4}C). Figure~\ref{cao_result_4}G shows the optimization curve, yielding an optimal correction coefficient of $\alpha^{\mathrm{corr}}_{0,2}=19.39$.

\section{\label{sec:level5}Conclusion}

In this work, we have proposed and experimentally demonstrated a sensorless AO framework based on the spatial coherence properties of light. By optimizing the localization of the measured spatial coherence distribution, the proposed framework restores the imaging system's PSF and compensates for optical aberrations. This eliminates the need for a physical guide star while remaining compatible with conventional incoherent illumination.

The proposed approach was validated under a range of experimentally relevant imaging conditions. Artificial phase elements were used to emulate system-induced and weak specimen-induced aberrations representative of practical microscopy systems. We further demonstrated robust aberration correction even in the presence of spatially structured background noise. In this work, while Zernike polynomials were used to represent aberrations, the proposed framework is equally compatible with other wavefront-shaping bases, such as Hadamard or random modes, which may be advantageous for more complex aberrations~\cite{hu2020aplp,yeminy2021scadv,vellekoop2007ol}.

Unlike the recently proposed intensity-intensity correlation approach~\cite{he2025aplph}, which requires specially engineered even-symmetrical thermal light, our framework operates with a standard LED source. 
Furthermore, our results show that the use of spatial correlations as the feedback metric for adaptive optics is not exclusive to entangled-photon illumination~\cite{cameron2024sc}, but can also be realized using the classical spatial correlations of incoherent light. This avoids the low photon-pair generation efficiency of entangled-photon sources and the complex, time-consuming, and inefficient coincidence measurements required for their detection. By using bright LED illumination and direct intensity measurements, our approach provides higher measurement photon flux and SNR and can be readily integrated into standard imaging systems.

A current limitation of the proposed method is the requirement for a wavefront-inversion interferometer, which demands accurate alignment and mechanical stability during CSD measurements.

The proposed framework relies only on the spatial incoherence of the illumination field and is therefore not restricted to the bright-field configuration demonstrated here. Since fluorescence emission is inherently spatially incoherent, the present approach is, in principle, compatible with fluorescence microscopy, potentially extending correlation-based AO to imaging modalities beyond those currently accessible with quantum implementations.

The present results demonstrate that spatial coherence measurements can serve as an effective feedback mechanism for sensorless AO. We anticipate that this framework will motivate further exploration of coherence-domain feedback strategies for different imaging systems.

\section{\label{sec:level1-Methods}Methods}

\subsection{Theoretical Framework and System Definition}

We consider a standard $4f$ imaging configuration operating under spatially incoherent illumination. The cross-spectral density (CSD) of such an idealized incoherent source at the sample plane is,
\begin{equation}
    W_{\mathrm{obj}}(\bm{\rho}_1, \bm{\rho}_2) = I_0(\bm{\rho}_1) \delta(\bm{\rho}_1 - \bm{\rho}_2),
\end{equation}
where $I_0(\bm{\rho})$ represents the spatial intensity distribution of the sample. 

The CSD at the image plane $W_{img}$ is found by propagating the source through the system impulse response $h(\mathbf{r})$, also referred to as the point spread function (PSF). For ease of calculation, assuming PSF to be shift-invariant, we get,
\begin{equation}
\scalebox{0.95}{$
\begin{aligned}
W_{img}(\mathbf{r}_1, \mathbf{r}_2)
&= \iint W_{obj}(\bm{\rho}_1, \bm{\rho}_2)
h(\mathbf{r}_1-\bm{\rho}_1)
h^*(\mathbf{r}_2-\bm{\rho}_2)
\, d\bm{\rho}_1 d\bm{\rho}_2
\\
&= \int I_0(\bm{\rho})
h(\mathbf{r}_1-\bm{\rho})
h^*(\mathbf{r}_2-\bm{\rho})
\, d\bm{\rho}
\end{aligned}
$}
\end{equation}
The coherent PSF, $h(\mathbf{r})$, is related via a Fourier transform to the complex pupil function $P(\mathbf{k}) = A(\mathbf{k}) e^{i\phi(\mathbf{k})}$, where $A(\mathbf{k})$ defines the physical pupil aperture and $\phi(\mathbf{k})$ represents the wavefront aberration phase profile.

To access the spatial coherence of the field, we employ a wavefront-inversion interferometer positioned at the exit plane. The inversion interferometer inherently measures the two-point correlation between $\mathbf{r}$ and $-\mathbf{r}$. Consequently, the measured inversion CSD simplifies to,
\begin{equation}
    W_{\mathrm{inv}}(\mathbf{r}) = W_{\mathrm{img}}(\mathbf{r}, -\mathbf{r}) = \int I_0(\bm{\rho}) h(\mathbf{r} - \bm{\rho}) h^*(-\mathbf{r} - \bm{\rho}) \, d\bm{\rho}.
\end{equation}
To guide our AO optimization loop, we define a coherence sharpness metric, $M$, corresponding to the $L_2$ norm of the inversion CSD. From Parseval's theorem, this optimization metric can be computed equivalently in either the spatial or spatial-frequency domain,
\begin{equation}
    M = \int |W_{\mathrm{inv}}(\mathbf{r})|^2 \, d\mathbf{r} = \int |\tilde{W}_{\mathrm{inv}}(\mathbf{u})|^2 \, d\mathbf{u},
\end{equation}
where $\tilde{W}_{\mathrm{inv}}(\mathbf{u})$ is the spatial Fourier transform of $W_{\mathrm{inv}}(\mathbf{r})$. By substituting the structural relationship between the PSF and the pupil function $P(\mathbf{k})$ into the integral, the spectral representation of the inversion CSD yields,
\begin{equation}
\begin{aligned}
    \tilde{W}_{\mathrm{inv}}(\mathbf{u}) = \frac{1}{4} \int \tilde{I}_0(\mathbf{v}) K(\mathbf{u}, \mathbf{v}) e^{i \Delta \phi(\mathbf{u}, \mathbf{v})} \, d\mathbf{v},
\end{aligned}
\end{equation}
where $\tilde{I}_0(\mathbf{v})$ denotes the Fourier transform of the sample intensity distribution, $K(\mathbf{u}, \mathbf{v}) = A\left(\frac{\mathbf{u}+\mathbf{v}}{2}\right)A\left(\frac{\mathbf{u}-\mathbf{v}}{2}\right)$ represents the overlapping pupil aperture support kernel, and $\Delta \phi(\mathbf{u}, \mathbf{v}) = \phi\left(\frac{\mathbf{u}+\mathbf{v}}{2}\right) - \phi\left(\frac{\mathbf{u}-\mathbf{v}}{2}\right)$ defines the phase difference profile.

\subsubsection{Metric response for Defocus aberration}
 We analyse a system experiencing defocus aberration under an unbounded clear aperture assumption ($A(\mathbf{k}) = 1$). The wavefront error is parameterized by the defocus phase coefficient $\alpha$:
\begin{equation}
    P(\mathbf{k}) = \exp(i \alpha |\mathbf{k}|^2).
\end{equation}
Evaluating the symmetric pupil function product at the split coordinate locations yields,
\begin{equation}
\scalebox{0.90}{$
\begin{aligned}
P\left(\frac{\mathbf{u}+\mathbf{v}}{2}\right)
P^*\left(\frac{\mathbf{u}-\mathbf{v}}{2}\right)
&=
\exp\left[
i \frac{\alpha}{4} |\mathbf{u}+\mathbf{v}|^2
\right]
\exp\left[
-i \frac{\alpha}{4} |\mathbf{u}-\mathbf{v}|^2
\right]
\\
&=
\exp\left[
i \alpha
(\mathbf{u}\cdot\mathbf{v})
\right]
\end{aligned}
$}
\end{equation}
Substituting this exact relation back into the inversion CSD spectral definition gives,
\begin{equation}
    \tilde{W}_{\mathrm{inv}}(\mathbf{u}) = \frac{1}{4} \int \tilde{I}_0(\mathbf{v}) \exp(i \alpha \mathbf{u}\cdot\mathbf{v}) \, d\mathbf{v}.
\end{equation}
This corresponds structurally to an inverse Fourier transform of the sample intensity spectrum evaluated at scaled spatial coordinates. Thus, it can be mapped directly back to the spatial domain distribution:
\begin{equation}
    \tilde{W}_{\mathrm{inv}}(\mathbf{u}) = \frac{1}{4} \, I_0\left( \frac{\alpha \mathbf{u}}{2\pi} \right).
\end{equation}
Calculating the total metric value $M(\alpha)$ through integration across the frequency space:
\begin{equation}
    M(\alpha) = \int |\tilde{W}_{\mathrm{inv}}(\mathbf{u})|^2 \, d\mathbf{u} = \frac{1}{16} \int \left| I_0\left( \frac{\alpha \mathbf{u}}{2\pi} \right) \right|^2 \, d\mathbf{u}.
\end{equation}
By applying a coordinate change of variables $\mathbf{u}' = \frac{\alpha \mathbf{u}}{2\pi}$, the metric evaluates to,
\begin{equation}
    M(\alpha) = \frac{\pi^2}{4\alpha^2} \int |I_0(\mathbf{u}')|^2 \, d\mathbf{u}'.
\end{equation}
The above equation demonstrates that the magnitude of the coherence metric $M(\alpha)$ scales inversely with the square of the defocus coefficient ($M \propto \alpha^{-2}$). Consequently, the metric monotonically decreases as the system drifts away from ideal focus, ensuring a global maximum at the perfectly corrected position ($\alpha = 0$).

\begin{figure*}[t]
\centering
\includegraphics[scale=0.74]{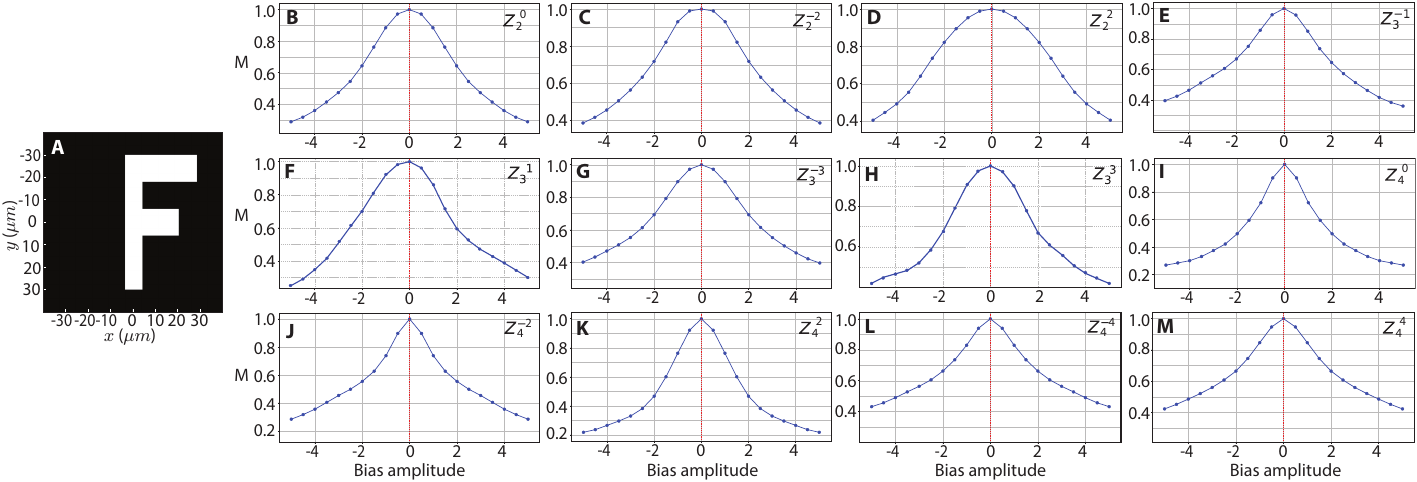}
\caption{Metric stability for a spatially asymmetric object. (A) The asymmetric object ("F") used for simulation. (B–M) Response curves of the coherence metric $M$ versus bias amplitude for individual Zernike modes ($2 \leq n \leq 4$, labeled on each plot). The monotonic decrease away from the origin confirms the metric's stability.}
\label{cao_asymm_metric}
\end{figure*}

\subsubsection{Arbitrary weak aberrations with symmetric object}
To confirm that the diffraction-limited state constitutes the maximum for optimization, we examine the behaviour of metric $M$ under a weak pupil phase aberration approximation ($\phi(\mathbf{k}) \ll 1$) across a circular aperture $A(\mathbf{k})$.

First, consider the mathematical parity of the object spectrum. For any physical spatial intensity distribution which is real-valued and non-negative, $I_0(\mathbf{r}) \in \mathbb{R}$, its Fourier transform $\tilde{I}_0(\mathbf{u})$ naturally exhibits Hermitian symmetry:
\begin{equation}
    \tilde{I}_0(\mathbf{u}) = \int_{-\infty}^{\infty} I_0(\mathbf{r}) e^{-i 2\pi \mathbf{u} \cdot \mathbf{r}} \, d\mathbf{r} \implies \tilde{I}_0(\mathbf{u}) = \tilde{I}_0^*(-\mathbf{u}).
\end{equation}
Decomposing $\tilde{I}_0(\mathbf{u})$ into its constituent real part $R(\mathbf{u})$ and imaginary part $X(\mathbf{u})$ yields
\begin{equation}
    R(\mathbf{u}) + iX(\mathbf{u}) = [R(-\mathbf{u}) + iX(-\mathbf{u})]^* = R(-\mathbf{u}) - iX(-\mathbf{u}),
\end{equation}
which establishes that the real spectral component is strictly even, $R(\mathbf{u}) = R(-\mathbf{u})$, while the imaginary component is strictly odd, $X(\mathbf{u}) = -X(-\mathbf{u})$.

Next, we evaluate the parity properties of the pupil-plane operations. The phase difference function $\Delta \phi(\mathbf{u}, \mathbf{v})$ is inherently odd with respect to the integration variable $\mathbf{v}$, regardless of the underlying aberration profile:
\begin{equation}
    \Delta \phi(\mathbf{u}, -\mathbf{v}) = \phi\left( \frac{\mathbf{u} - \mathbf{v}}{2} \right) - \phi\left( \frac{\mathbf{u} + \mathbf{v}}{2} \right) = - \Delta \phi(\mathbf{u}, \mathbf{v}).
\end{equation}
Concurrently, the circular symmetry of the physical pupil aperture ensures that the overlapping support kernel remains strictly even with respect to $\mathbf{v}$, such that $K(\mathbf{u}, \mathbf{v}) = K(\mathbf{u}, -\mathbf{v})$.

Considering the weak aberration assumption, we expand the complex phase exponential via a Taylor series up to the second order, $e^{i\Delta\phi} \approx 1 + i\Delta\phi - \frac{1}{2}\Delta\phi^2$. The spectral representation can be written as
\begin{equation}
    \tilde{W}_{\mathrm{inv}}(\mathbf{u}) \approx \frac{1}{4} \int [R(\mathbf{v}) + iX(\mathbf{v})] K(\mathbf{u}, \mathbf{v}) \left[1 + i\Delta\phi - \frac{1}{2}\Delta\phi^2\right] \, d\mathbf{v}
\end{equation}
Evaluating the integrands over symmetric limits based on their parity with respect to $\mathbf{v}$ gives,
\begin{equation}
    \tilde{W}_{\mathrm{inv}}(\mathbf{u}) \approx \frac{1}{4} \left[ S_K(\mathbf{u}) - \int X K \Delta \phi \, d\mathbf{v} - \frac{1}{2} \int R K \Delta \phi^2 \, d\mathbf{v} \right].
\end{equation}
By defining the simplified integral terms $\mathcal{X} = \int X K \Delta \phi \, d\mathbf{v}$ and $\mathcal{R} = \int R K \Delta \phi^2 \, d\mathbf{v}$, we evaluate the squared magnitude of the spectrum, discarding terms higher than $O(\phi^2)$:
\begin{equation}
\begin{aligned}
|\tilde{W}_{inv}|^2 &= \frac{1}{16} [S_K - \mathcal{X} - \frac{1}{2}\mathcal{R}]^2 \\ &\approx \frac{1}{16} [S_K^2 - 2 S_K \mathcal{X} + \mathcal{X}^2 - S_K \mathcal{R}]
\end{aligned}
\end{equation}
For a spatially symmetric sample distribution, $I_0(\mathbf{r}) = I_0(-\mathbf{r})$, the imaginary spectral component vanishes entirely ($X(\mathbf{u}) = 0$) and this equation reduces directly to
\begin{equation}
    |\tilde{W}_{\mathrm{inv}}|^2 \approx \frac{1}{16} \left[S_K^2  - S_K \mathcal{R}\right].
\end{equation}
Integrating over the spatial frequency coordinates $\mathbf{u}$, the net change in the optimization metric $\Delta M$ resulting from the introduction of phase distortions is given by
\begin{equation}
    \Delta M \approx -\frac{1}{16} \int_{\mathbf{u}}  \left( \int R K \, d\mathbf{v} \right) \left( \int R K \Delta \phi^2 \, d\mathbf{v} \right) \, d\mathbf{u}.
\end{equation}
For weak aberration, $\Delta M$ is negative. So, the metric decreases when aberration is introduced.

\subsubsection{Numerical validation for spatially asymmetric object}

While the analytical formulation derived above formally confirms metric stability for centrosymmetric object distributions, real-world label-free microscopy samples rarely exhibit perfect spatial symmetry. To evaluate the resilience of the coherence sharpness metric $M$ under general, asymmetric conditions, we perform numerical simulations using a highly asymmetric target profile. 

We choose the capital letter ``F'' as our test sample [see Fig.~\ref{cao_asymm_metric}(A)], since its geometry lacks any standard orthogonal or radial reflection symmetry, meaning its imaginary spectral component $X(\mathbf{u})$ remains non-zero. To model realistic physical aberrations, the imaging system's pupil plane is perturbed with various low-order Zernike modal distortions spanning radial orders $2 \leq n \leq 4$. The response of the coherence metric $M$ is then systematically tracked as a function of the applied aberration bias amplitude.

The results of these numerical investigations are presented in Figs.~\ref{cao_asymm_metric}(B--M). Crucially, across all tested Zernike modes, the measured metric decreases as the aberration magnitude increases from its optimal value. These findings demonstrate that for a general object subjected to weak, slowly varying wavefront errors, the maximum of the coherence sharpness metric remains at the aberration-free state ($\phi = 0$). This numerical validation confirms that the spatial-coherence feedback mechanism provides a stable optimization metric for arbitrary sample structures.

\subsection{\label{sec:level1}Cross-spectral denisty measurement}

To experimentally measure the cross-spectral density function, we implement a phase-shifting wavefront-inversion interferometric system. It works by interfering the optical field with a spatially inverted replica of itself with controlled phase shifts between the two interferometer arms. Recording interferograms at four distinct phase offsets enables the retrieval of both the real and imaginary components of the CSD.

As shown in Fig.~\ref{cao_setup}, the measurement system is based on a Mach-Zehnder interferometer geometry. Wavefront inversion in the two arms is implemented using Dove prisms. Since Dove prisms may introduce corner blockage and unwanted front-surface reflections, K-mirror assemblies can alternatively be employed, as discussed in Ref.~\cite{karan2022ao}. Phase modulation between the interferometer arms is achieved using a geometric-phase unit composed of two quarter-wave plates oriented at \(45^\circ\) with respect to the input polarizer and a half-wave plate (HWP) inserted between them in one interferometer arm. Rotating the HWP by an angle \(\theta\) introduces a geometric phase delay of \(2\theta\) between the interfering fields~\cite{jha2008prl}.

The intensity distribution recorded at the camera plane can be expressed as,
\begin{small}
\begin{align}
I_{\mathrm{out}}(\mathbf{r}; \delta) = & \ k_1^2 I_1(-\mathbf{r}) + k_2^2 I_2(\mathbf{r}) \nonumber  + 2k_1 k_2 \mathrm{Re}\left[W_{\mathrm{img}}(-\mathbf{r}, \mathbf{r})\right] \cos \delta \nonumber  - 2k_1 k_2 \mathrm{Im}\left[W_{\mathrm{img}}(-\mathbf{r}, \mathbf{r})\right] \sin \delta,
\label{intensity_out}
\end{align}
\end{small}
where $\delta = \beta_2 - \beta_1$ denotes the phase difference between the two interferometer arms, $I_1(-\mathbf{r})$ and $I_2(\mathbf{r})$ represent the individual arm intensities, and $W_{\mathrm{img}}(-\mathbf{r}, \mathbf{r})$ is the cross-spectral density function evaluated at the inverted coordinate pair. The symbols \(\text{Re}[\cdots]\) and \(\text{Im}[\cdots]\) denote the real and imaginary parts of the CSD, respectively.

To extract the CSD, the output intensity is recorded for multiple values of the phase difference \(\delta\). Considering two phase settings, \(\delta_c\) and \(\delta_d\), the corresponding difference intensity is given by,
\begin{equation}
\Delta I^{(\delta_c, \delta_d)}_{\mathrm{out}}(\mathbf{r}) = 2k_1k_2 \Big\{ \mathrm{Re}\left[W_{\mathrm{img}}(-\mathbf{r}, \mathbf{r})\right] (\cos\delta_c - \cos\delta_d)
- \mathrm{Im}\left[W_{\mathrm{img}}(-\mathbf{r}, \mathbf{r})\right] (\sin\delta_c - \sin\delta_d) \Big\}.
\label{interferometer_intensity_diff}
\end{equation}
Choosing \(\delta_c = 0\) and \(\delta_d = \pi\) isolates the real component of the CSD,
\begin{equation}
    \mathrm{Re}\left[W_{\mathrm{img}}(-\mathbf{r}, \mathbf{r})\right] = \frac{\Delta I_{\mathrm{out}}^{(0, \pi)}(\mathbf{r})}{4k_1 k_2}.
\label{ccf_real}
\end{equation}
Similarly, setting \(\delta_c = 3\pi/2\) and \(\delta_d = \pi/2\) yields the imaginary component:
\begin{equation}
    \mathrm{Im}\left[W_{\mathrm{img}}(-\mathbf{r}, \mathbf{r})\right] = \frac{\Delta I_{\mathrm{out}}^{(3\pi/2, \pi/2)}(\mathbf{r})}{4k_1 k_2}.
\label{ccf_imag}
\end{equation}
Thus, by measuring the interferometric intensity differences at these phase settings, both the real and imaginary parts of the cross-spectral density function can be reconstructed up to a common multiplicative factor of \(4k_1k_2\). This approach avoids the need for independent calibration of the parameters \(k_1\), \(k_2\), \(I_1\), and \(I_2\).

\section{\label{sec:level1}Acknowledgement}
We acknowledge financial support from the Science and Engineering Research Board through grants STR/2021/000035 and CRG/2022/003070, from the Department of Science and Technology, Government of India, through grant DST/ICPS/QuST/Theme-1/2019 and through the National Quantum Mission (NQM) technical group (TG) project on quantum imaging. H.D. acknowledges funding from ERC starting grant SQIMIC-101039375. P.M. thanks the Prime Minister Research Fellowship (PMRF), Government of India, for financial support.

\bibliography{EncDec_ref}    

\end{document}